\documentclass[a4paper,fleqn]{cas-dc}

\usepackage[round,authoryear]{natbib}

\usepackage{subfigure}
\usepackage{times}
\usepackage{hyperref}
\usepackage[ruled]{algorithm2e}
\usepackage{soul}
\usepackage{amsmath}
\usepackage{amsfonts}
\usepackage{amssymb}

\usepackage{afterpage}
\usepackage{bm}
\usepackage{multirow}
\usepackage{graphicx}
\usepackage{color, xcolor}

\soulregister\cite7
\soulregister\citep7
\soulregister\ref7
 
\def\tsc#1{\csdef{#1}{\textsc{\lowercase{#1}}\xspace}}
\tsc{WGM}
\tsc{QE}

\begin{document}
\let\WriteBookmarks\relax
\def\floatpagepagefraction{1}
\def\textpagefraction{.001}

\begin{titlepage}
    \vspace*{\fill}
    \begin{center}
        \normalfont{Multi-behavior Recommendation with SVD Graph Neural Networks}

        \vspace{2cm}

        \bigskip
        {Shengxi Fu$^a$, Qianqian Ren$^{a,*}$, Xingfeng Lv$^a$ and Jinbao Li$^{b,*}$}

        \bigskip
        {$^a$Department of Computer Science and Technology, Heilongjiang University, Harbin 150080, China\\
        $^b$Shandong Artificial Intelligence Institute,School of Mathematics and Statistics, Qilu University of Technology, Jinan 250014, China}
       
       \bigskip
        {$^*$ Corresponding author: Qianqian Ren}

        \bigskip
        {E-mail address: 2221912@s.hlju.edu.cn, renqianqian@hlju.edu.cn, lvxingfeng@hlju.edu.cn, lijinb@sdas.org}
        \medskip
    \end{center}
    \vspace*{\stretch{3}}
\end{titlepage}




\begin{abstract}
Graph Neural Networks (GNNs) have been extensively employed in the field of recommendation systems, offering users personalized recommendations and yielding remarkable outcomes. Recently, GNNs incorporating contrastive learning have demonstrated promising performance in handling the sparse data problem of recommendation systems. However, existing contrastive learning methods still have limitations in resisting noise interference, especially for multi-behavior recommendation. To mitigate the aforementioned issues, this paper proposes a GNN-based multi-behavior recommendation model called MB-SVD that utilizes Singular Value Decomposition (SVD) graphs to enhance model performance. In particular, MB-SVD considers user preferences across different behaviors, improving recommendation effectiveness. First, MB-SVD integrates the representation of users and items under different behaviors with learnable weight scores, which efficiently considers the influence of different behaviors. Then, MB-SVD generates augmented graph representation with global collaborative relations. Next, we simplify the contrastive learning framework by directly contrasting original representation with the enhanced representation using the InfoNCE loss. 
Through extensive experimentation, the remarkable performance of our proposed MB-SVD approach in multi-behavior recommendation endeavors across diverse real-world datasets is exhibited. 
\end{abstract}
\maketitle

\section{Introduction}
\begin{figure}[ht]
\centering
\includegraphics[width=0.95\linewidth]{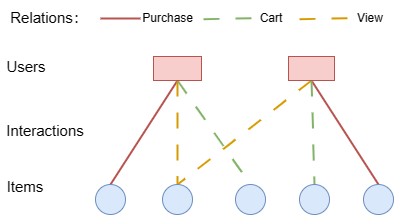}
  \caption{\rm{Scenarios of multiple behavior recommendations in e-commerce. It illustrates the key elements of the scenarios, including users, items, and various types of behaviors. The red line represents purchase behavior, the green line indicates view behavior, and the yellow line represents add-to-cart behavior.}}
\label{t1}
\end{figure}

Given the exponential growth of information, personalized recommendation systems have emerged as a crucial technology in many online services, which helps users filtering personalized content and improving user experience \citep{he2020lightgcn,mao2021ultragcn,peng2022svd}. Many efforts have been devoted to provide efficient recommendation. However, most existing recommendation methods are predicated on the foundation of individual behavioral patterns, also known as single-behavior recommendation models. Recently, more and more studies have found that single behavior data is no longer sufficient to meet the requirements of recommendation systems \citep{xia2021graph,wu2022multiview,huang2021graph}, while multiple types of behavior data has great potential to help predicting the interests and  behaviors of users \citep{8731537}. 
Let's take the scene depicted in Fig. \ref{t1} as an example.
In real-world scenarios, user behavior in e-commerce is typically multi-faceted and consists of more than one type of interaction. In addition to purchase behavior, factors such as view histories and add-to-cart behavior are also considered as valuable indicators of user preferences. By considering multiple types of behaviors, the recommendation system can gain a more comprehensive understanding of user preferences and provide more accurate and personalized recommendations. 

In recent years, multi-behavior recommendation models have attracted many interests and attentions. These models comprehensively consider various user behaviors to better understand their preferences and requirements, which efficiently enhances the accuracy and effectiveness of recommendation systems \citep{8731537,jin2020multi}. In particular, graph neural networks(GNNs) have been employed for multi-behavior modeling and feature extraction.
For the study of multi-behavior recommendation, weighting different behaviors of users to decide their different influence is a key problem. Some studies have treated the correlations among users and behaviors as the graph problem. Moreover, they design scheme to assign trainable weights to edges of the graph, and capture the significance of individual behaviors \citep{xia2021knowledge}. Constructing embedding representation for each type of behavior is also an efficient solution, which  combines node embeddings with graph evolution operations \citep{chengraph}. By combining behavior embeddings and node embeddings, we can better understand behaviors and interests of users, thus improve recommendation effectiveness. Recently, the commonalities and differences between target behavior and other behaviors are explored. Incorporating the shared characteristics of disparate behaviors into the target behavior facilitates the enhancement of its embedding representation\citep{gu2022self}.

Although incorporating GNNs into multi-behavior recommendation have achieved certain results in improving recommendation effectiveness, however, two significant challenges still remain unresolved.
\begin{itemize}
    \item \textbf{Over-Smoothing Problem in Deeper GNNs.} It usually suffers from over-smoothing problem when processing the correlations among multiple behaviors with deeper GNNs. This over-smoothing feature may reduce the embedding quality of target behaviors, thereby affecting the accuracy and effectiveness of the recommendation system. 
\item \textbf{Data Noise and Sparsity.} In real-world scenarios, the performance of recommendation systems is significantly influenced by the presence of noise and data sparsity. These factors pose challenges to the effectiveness of behavior embedding techniques, which are essential for capturing user preferences and generating accurate recommendation.
\end{itemize} 

Therefore, mitigating the issue of over-smoothing within models and enhancing the inherent embedding representation of target behaviors are the key problems to be addressed. In this paper, we introduce MB-SVD, a novel multi-behavior recommendation model that combines the advantages of singular value decomposition (SVD) techniques and graph convolutional networks (GCN). In our approach, we focus on learning distinct embedding vectors for each behavior, allowing us to capture the unique characteristics of different behaviors. To combine these representations effectively, we introduce a behavior-aware aggregation model that automatically learns weights to integrate the representations of multiple behaviors. This aggregation process ensures that the combined representation captures the essential features from each behavior. 
Moreover, we propose SVD augmentation strategy to enhance the aggregated representations. This strategy efficiently preserves important behavior features from both the specific user and cross users. By incorporating SVD augmentation, we can enrich the representations with additional information, thus improve their quality and robustness. To address the challenges of data noise and sparsity, we design a simplified contrastive learning paradigm, which compares the original user and item representations with the SVD augmented representations.

The summarization of our contributions is outlined as follows:

\begin{itemize}
\item In this paper, we enhance the multi-behavior recommendation models by proposing a novel simplified graph contrastive learning framework to improve the model performance and address the data sparsity and noise problem simultaneously.

\item We design an effective contrastive learning paradigm named MB-SVD, which constructs contrastive view pairs with the SVD-augmentation representation view and the original multi-behaviors weighted representation. It is efficient to distill important collaborative signals from the global perspective.

\item  We conduct extensive experiments on three real-world datasets, and the results demonstrate that our model has better recommendation performance compared to other baseline models.
\end{itemize}

The organization of the paper is arranged as follows. Section 2 discusses related work. Section 3 gives the problem formulation. Section 4 details the proposed model. Section 5 presents an extensive experimental evaluation. Section 6 concludes the paper.

\section{Related Work}
Recommendation system has attracted continuous attentions and many efficient solutions have been proposed.
Recently, Graph Neural Networks (GNNs) are widely used in recommendation systems \citep{gao2021graph}. 
By leveraging its capability to aggregate user-item similarities and  leverage the structural connectivity of the graph, GNNs exhibit great power to generate excellent user and item embeddings, thereby enhancing the quality of recommendations. NGCF \citep{wang2019neural} introduces a spatial GNN paradigm for recommendation tasks, exhibiting superior performance compared to conventional collaborative filtering (CF) approaches.
LightGCN \citep{he2020lightgcn} leverages an embedding learning approach for users and items representation, employing a weighted aggregation of embeddings across multiple layers to generate the integrated representation.
SGL\citep{SGL} is proposed for improving recommendation systems and enhancing node representation learning by self-discrimination.
GCA\citep{GCA} incorporates user and item features, along with user-item interactions with a graph convolution network and a gated fusion mechanism.
HCCF\citep{HCCF} introduces hyper-graph structure learning that enhances the discrimination ability of GNN-based CF paradigm, which comprehensively captures the complex high-order dependencies among users.
With the distilled global context, SHT\citep{SHT} employs a cross-view generative self-supervised learning component for data augmentation over the user-item interaction graph, enhancing the robustness of recommendation systems.
SimGCL\citep{SimGCL}  boosts the representation quality of recommendation system by strengthening self-discrimination, which improves the robustness of recommendation systems.

In recent years, multi-behavior recommendation has emerged as a novel area in recommendation system. Different with single behavior recommendation models, multi-behavior recommendation models integrate multiple different behaviors to improve the precision of user-targeted behavior recommendation \citep{chen2020efficient, wang2021self}. B-GMN \citep{xia2021graph} enhances the learning of user-item interactions by automatically distilling behavior heterogeneity and interaction diversity for improved recommendation. MATN\citep{MATN} designs an attention network for multi-behavior recommendation, which learns user and item embeddings from multiple behavior data with attention mechanisms to capture behavior importance and correlations. MBGCN\citep{MBGCN} is a recommendation model that leverages graph convolution networks to incorporate user and item features, as well as user-item interactions. It employs gated fusion and residual learning techniques to enhance its performance.
EHCF\citep{EHCF} proposes a cutting-edge approach in collaborative filtering for heterogeneous data, which is efficient to capture fine-grained interactions between users and items.
NMTR\citep{NMTR} is an advanced multi-faceted recommendation model that operates across multiple tasks, effectively leveraging diverse behavior data for collaborative filtering. 
S-MBRec\citep{gu2022self} focuses on leveraging the connectivity and inherent differences between the target behavior and auxiliary behaviors to improve recommendation performance.

In summering existing multi-behavior recommendation methods, it is evident that the transition smoothing issue arising from the convolution of multi-task behavior has not been adequately addressed. Moreover, the problems of noise and data sparsity persist despite the numerous attempts to tackle them.

\begin{figure*}

\centering\includegraphics[width=\linewidth,height=0.62\textwidth]{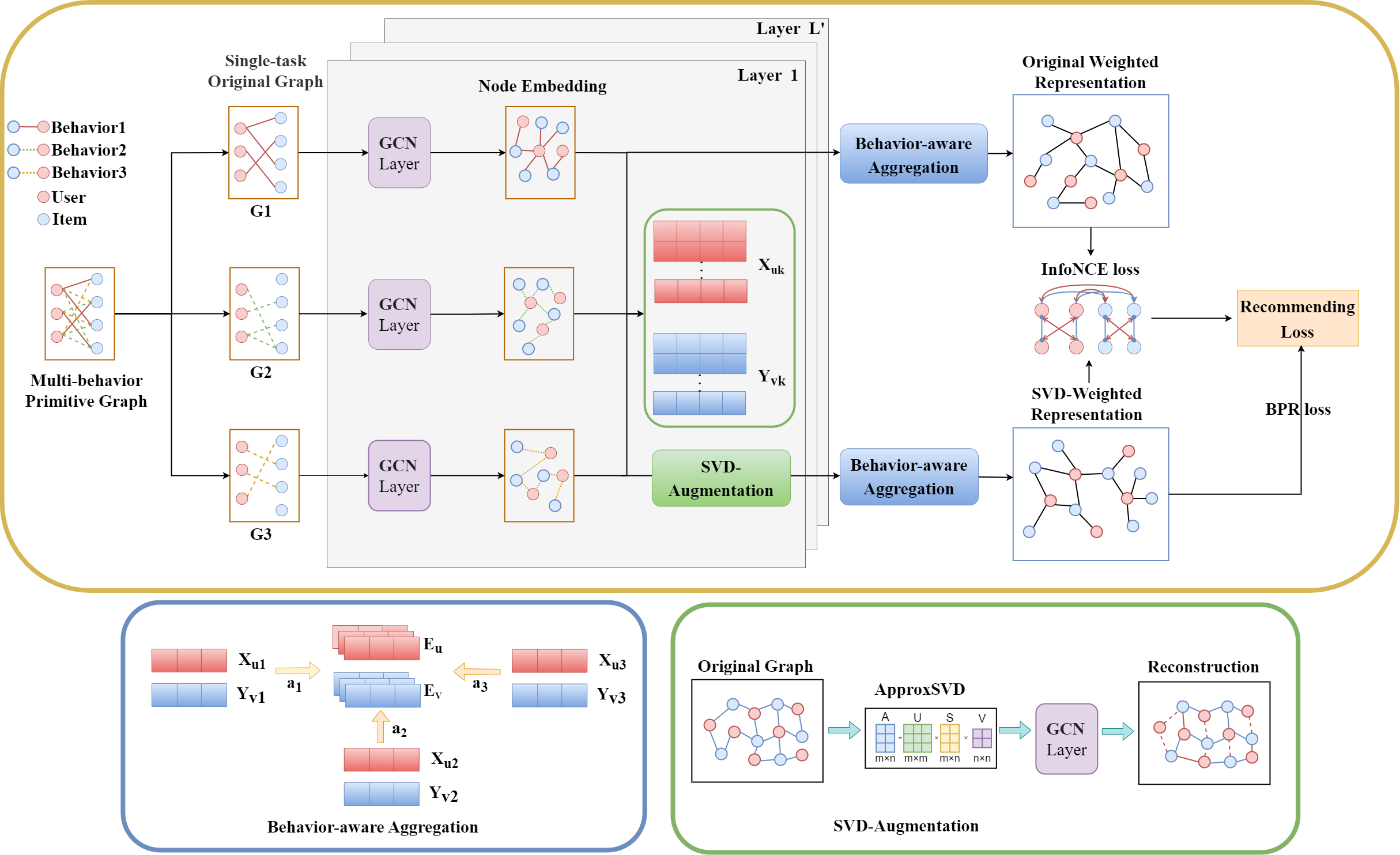}
   \caption{The architectural framework of MB-SVD. We illustrate the model with an instance where $K$ is equal to 3, indicating that there are three types of behaviors.}
   \label{architecture}
 \end{figure*}

\section{Problem Statement}\label{3}
In this section, we formalize the problem of multi-behavior recommendation. First, we provide some relevant definitions.

$\textbf{Definition 1.}$ (Graph). A graph can be represented by ${G} = ({V,E})$, where ${V}=\{v_j\}_{j=1,2,\cdots,N}$ contains a group of nodes, and  ${E}=\{e_{ij}\}$ contains a group of connections. In particular, we distinguish two types of nodes in $V$ , user nodes $u \in U$ and item nodes $i \in I$, connections in $E$ that reflect user-item interactions.

$\textbf{Definition 2.}$ ($k$-behavior Graph) A $k$-behavior graph consists of $G_k = (V, E_k, A_k)$, where $k=\{1,2,\cdots K\}$, $K (K \geq 2)$ indicates the number of distinct user-item behaviors, and $G_k\in G$. The edge set for the $k$-{th} behavior is indicated by $E_k$. 
$A_k \in R^{|U|*|I|}$ represents the interaction matrix of the graph under $k-$th behavior. To simplify notation, we will use $A$ instead in the remaining sections of the paper.

$\textbf{Multi-behavior Recommendation Problem}$. 
Given various types of behavior data, multi-behavior recommendation aims to predict the possibility of interaction between users and items.

\section{METHODOLOGY}
 This section elaborates on the proposed MB-SVD framework and its various components design.
 \subsection{Overview}
Figure \ref{architecture} depicts the overall architecture of our MB-SVD model, which is composed of two phases, node representation and co-contrasting learning. For the given multi-behavior graph, we begin by partitioning it into multiple subgraphs. Then, we utilize GCNs to obtain the node embeddings for each behavior graph. In the co-contrastive learning phase,  
to discriminate the salience of different behaviors and capture the differences among behavior embeddings, we propose a novel augmentation module aimed at autonomously acquiring weights, thereby boosting the learning process. Specifically, we utilize SVD to reconstruct the node embeddings to capture the global collaborative signals and patterns across behaviors. Next, in multi-behavior contrastive learning, we address the issues of smooth transition between different behavior types and sparse data by performing contrastive learning on two types of embeddings: the weighted fused node embeddings and the weighted fused node embeddings after SVD reconstruction. 
Finally, we fuse the embeddings of each behavior and jointly optimize the task.

\subsection{Behavior-guided Graph Representation}
The behavior-guided graph representation module aims to capture the behavior-specific patterns and features in user-item interactions. It takes into account the diverse behavioral contexts and their impact on user preferences. 
In particular, we first split behaviors of $K$-behavior graph to obtain $K$ sub-graphs of distinct behaviors, denoted as $\{G_1,G_2, \cdots,G_K\}$.
Specially, for each user $u$ and each item $v$, we assign embedding vectors $X_{uk}$ and $Y_{vk}$ of dimension $d$ at $k$-th behavior, where $k$ represents the behavior type
($k=1,2,\cdots,K$). 
The embedding vectors of all users and items at $k$-th behavior form $X_{k} \in R^{|U|\times d}$ and $Y_{k} \in R^{|I|\times d}$. 
To incorporate the information from neighboring nodes and preserve the original information of nodes, we employ a two-layer GCN along with residual connections \citep{xia2022hypergraph}. The update process for aggregated embeddings at $l-$th layer can be mathematically represented:
\begin{equation}
\begin{split}
x_{uk}^{(l)}&=\sigma(p(\widetilde{A}_{u,:})\cdot Y_{k}^{(l-1)})\\
y_{vk}^{(l)}&=\sigma(p(\widetilde{A}_{:,v})\cdot X_{k}^{(l-1)})\\
X_{uk}^{(l)}&=x_{uk}^{(l)}+X_{uk}^{(l-1)}\\
Y_{vk}^{(l)}&=y_{vk}^{(l)}+Y_{vk}^{(l-1)}
\end{split}
\end{equation}
where $x_{uk}^{(l)}$ and $y_{vk}^{(l)}$ refer to the aggregated embeddings of $l$-th layer for user $u$ and item $v$ at $k-$th behavior, respectively. $X_{k}^{(l-1)}$ and $Y_{k}^{(l-1)}$ represents the embeddings of the previous $(l-1)$-th layer for users and items at the $k$-th behavior, respectively. $\widetilde{A}_{u,:}$ and $\widetilde{A}_{:,v}$ represent the row and column vectors of the normalized adjacency matrix $A$. 
$\sigma(\cdot)$ is the ReLU activation function, and $p(\cdot)$ denotes edge dropout.

By accumulating its embeddings across all layers, we obtain the final representation of user $u$ and item $v$ at the $k$-th behavior as:

For user $u$:
\begin{equation}\quad\quad
X_{uk}=\sum_{l=0}^LX_{uk}^{(l)}
\end{equation}

For item $v$:
\begin{equation}\quad\quad
Y_{vk}=\sum_{l=0}^LY_{vk}^{(l)}
\end{equation}
where $L$ denotes the total number of layers in the GCN.

\subsection{Behavior-aware Aggregation}
After performing node representation and obtaining the embeddings for users and items, the model recognizes the different behaviors and obtains corresponding embedding representations for each behavior. 
To consider the influence of the number and impact intensity of each behavior, the model adaptively assigns different importance  to each behavior. In this section, a behavior-aware augmentation module is introduced that dynamically learns weight scores for node representations associated with diverse behavioral contexts. Therefore, the model can effectively capture the varying influence of different behaviors on the recommendation task. 

To enhance the semantic fusion between user $u$ and item $v$, we employ specific coefficient $a_k$ to measure the $k$-{th} behavior. $a_k$ accounts for both the frequency and intensity of $k$-th behavior, resulting in a more accurate representation of their correlations, that's
\begin{equation}\quad\quad\quad
a_k=\frac{exp(w_k * n_{uk})}{\sum_{m=1}^Kexp(w_m * n_{um})}
\end{equation}
where $w_k$ represents the uniform weighting factor for the $k$-th behavior, which is consistently applied across all users. It captures the significance of the behavior type on the user representations.
On the other hand, $n_{uk}$ quantizes the correlations of user $u$ within the $k$-th behavioral context. By using the combination of the $w_k$ and $n_{uk}$, the model can capture the influence of different behavior types on user representations. Moreover, it enables a more accurate understanding of user preferences and behavior patterns in the recommendation process.

Given the embedding matrices $X_{uk}$ and $Y_{vk}$ as the features of user $u$ and item $v$ within the $k$-{th} behavioral context,  we further integrate all representations across behaviors to obtain the final feature representation for user $u$ and item $v$: 

\begin{equation}\quad\quad\quad
\begin{split}
    E_u=\sigma \left(W_u(\sum_{k=1}^Ka_{k}*X_{uk})+b_u \right)\\
    E_v=\sigma \left(W_v(\sum_{k=1}^Ka_{k}*Y_{vk})+b_v \right)
\end{split}
\end{equation}
where $W_u$ and $W_v$ are the weight matrices that transform the aggregated representations of user $u$ and item $v$, respectively.
$b_u$ and $b_v$ are the bias terms associated with user $u$ and item $v$, respectively.

\subsection{Capturing Global Collaborative Dependencies with SVD}
To enable global structure learning for graph contrastive recommendation, inspired by \citep{rajwade2012image,rangarajan2001learning}, we further design global features extraction  model with SVD scheme. This scheme efficiently extracts essential collaborative signals from a global perspective. First, we reconstruct the adjacency matrix $A$ via performing SVD operator followed by truncation process as \citep{cai2023lightgcl}:
\begin{equation}\quad\quad\quad
\begin{split}
    &A = USV^T\\
    &\hat{A} = U_qS_qV_q^T
\end{split}
\end{equation}
where $U$ and $V$ are orthonormal matrices, their columns represent the eigenvectors of A's row-row correlation matrix and column-column correlation matrix, respectively. $S$ is a diagonal matrix that stores the singular values of $A$. 
$S_q$ is a diagonal matrix obtained by selecting only the top $q$ singular values from $S$ and setting the rest to zero. $U_q$ and $V_q$ are truncated matrix containing the first $q$ columns of $U$ and $V$, respectively. $\hat{A}$ is a low-dimensional approximation of $A$ with the target rank $q$.

According to \citep{cai2023lightgcl}, computing the exact SVD on large matrices, especially for vast user-item matrices, can be computationally expensive and memory-intensive, making it infeasible in practice. Instead, we solve the problem by utilizing  the random singular value decomposition technique \citep{halko2011finding}.
By approximating the matrix's range with a low-rank orthonormal matrix before performing the SVD, it significantly reduces the computational complexity compared to computing the exact SVD on the original large-scale matrix. The procedure is formulated as:
\begin{equation}\quad\quad\quad
\begin{split}
\hat{U_q},\hat{S_q},\hat{V_q^\top}&=\mathrm{f}(A,q)\\ 
\end{split}
\end{equation}
where $\mathrm{f}$ is the function that approximates the singular value decomposition of matrix $A$ with a target rank $q$, defined as \citep{cai2023lightgcl}.  $\hat{U}_q$, $\hat{S}_q$, and $\hat{V}_q$ are the approximated matrices of $U_q$, $S_q$, and $V_q$, respectively. The reconstructed matrix $\hat{A}_\mathrm{f}$ is obtained by multiplying the truncated matrices. Thus, we compute the user and item embeddings as a combination of the approximated matrices and the collective representations of the embeddings as follows:

\begin{equation}\quad
\begin{split}
 G_{uk}^{(l)}=\sigma (\hat{A}_\mathrm{f}Y_{k}^{(l-1)})=\sigma (\hat{U_q}\hat{S_q}\hat{V_q^\top}Y_{k}^{(l-1)})\\
 G_{vk}^{(l)}=\sigma  (\hat{A}_\mathrm{f}X_{k}^{(l-1)})=\sigma (\hat{U_q}\hat{S_q}\hat{V_q^\top}X_{k}^{(l-1)})
\end{split}
\end{equation}
where $G_{uk}^{(l)}$ and $G_{vk}^{(l)}$ denote user and item embedding sets at $k$-th behavior derived from the new constructed views. 
Upon applying the SVD on these embeddings, we obtain enhanced features for users and items, denoted as $G_{uk}$ and $G_{vk}$, respectively.

For user $u$:
\begin{equation}\quad\quad
G_{uk}=\sum_{l=0}^LG_{uk}^{(l)}
\end{equation}

For item $v$:
\begin{equation}\quad\quad
G_{vk}=\sum_{l=0}^LG_{vk}^{(l)}
\end{equation}

To represent the enhanced weighted fusion results of each user and item across $K$ behaviors, we can express it as follows:

\begin{equation}\quad\quad\quad
\begin{split} 
F_u=\sigma \left(W_u(\sum_{k=1}^Ka_{k}*G_{uk})+b_u \right)\\
F_v=\sigma \left(W_v(\sum_{k=1}^Ka_{k}*G_{vk})+b_v \right)
\end{split}
\end{equation}

Here, $\alpha_k$ represents the weight assigned to the enhanced feature $G_{uk}$ or $G_{vk}$ corresponding to the $k$-th behavior. 
The weighted fusion results, $F_{u}$ and $F_{v}$, represent the combined and weighted representations of users and items across the $K$ behaviors. These fusion results model the overall features of users and items by considering the enhanced features derived from the SVD operation of the individual behavior-specific embedding sets.

\subsection{Multi-behavior Contrastive Learning}
In this section, we introduce the multi-behavior contrastive learning task. In contrast to most existing methods that utilizes two additional views for contrasting node embeddings, we design a simplified multi-behavior contrastive learning paradigm inspired by the work of \citep{cai2023lightgcl}. In our approach, we directly compare the InfoNCE loss \citep{van2018representation} between the weighted SVD-augmented embedding $F_{u}^{(l)}$/$F_{v}^{(l)}$ and the weighted orignial multi-behavior representation embedding $E_{u}^{(l)}$/$E_{v}^{(l)}$.

The contrastive loss for user embeddings, denoted as $L_{us}$, and item embeddings, denoted as $L_{vs}$, can be defined as:
\begin{equation}
\begin{split}
L_{us}=\sum_{u=0}^M\sum_{l=0}^L-log\frac {exp\left(s(F_{u}^{(l)},E_{u}^{(l)}/\tau)\right)} 
{\sum_{u=0}^Mexp\left(s(F_{u}^{(l)},E_{u}^{(l)}/\tau)\right)}\\
L_{vs}=\sum_{v=0}^N\sum_{l=0}^L-log\frac {exp\left(s(F_{v}^{(l)},E_{v}^{(l)}/\tau)\right)} {\sum_{v=0}^Nexp\left(s(F_{v}^{(l)},E_{v}^{(l)}/\tau)\right)}
\end{split}
\end{equation}

The cosine similarity function $s(.)$ and the temperature parameter $\tau$ are incorporated in the InfoNCE loss $L_{us}$ and $L_{vs}$ for users and items, respectively. 

\subsection{Loss Function and Joint Optimization}
To enhance the computational efficiency of the model and optimize the contrastive loss, we further introduce an advanced formulation of the Paired Bayesian Personalized Ranking (BPR) loss function\citep{rendle2012bpr}. This enhanced formulation emphasizes the similarities among connected nodes, which can lead to better performance in recommendation systems. Specifically, we construct the BPR loss function as follows:
\begin{equation}\quad\quad
L_r=\sum_{(u,i,j)\in O}-log\left(\sigma(F_u^\top F_i-F_u^\top F_j)\right)
\end{equation}

Given the training dataset  $O =\{(u, i, j)|(u, i) \in O_+, (u, j) \in O_- \}$, where $O_+$ represents the collection of observed interactions, and $O_-$ is characterized as the collection of all unobserved interactions.
This formulation emphasizes the importance of observed interactions while considering the complementary set of unobserved interactions, thereby enabling the model to capture the preferences and patterns of the users  based on the entire item space and generalize its predictions. It leads to more effective and accurate recommendations.

Next, the contrastive loss is jointly optimized with the main objective function for the multi-behavior recommendation task. The overall objective function $L$ is defined as:
\begin{equation}\quad\quad
L=L_r+\lambda(L_{us}+L_{vs})+\beta||\Theta||_2^2
\end{equation}
where $\lambda$ is a hyperparameter that controls the importance given to the contrastive loss relative to the main objective function. 
$\beta ||\Theta||_2^2$ introduces $L_2$ regularization to the objective function. Here, $\Theta$ denotes the complete set of trainable parameters within the model. By adding this term, the objective function is regularized to prevent overfitting and encourage generalization.

\section{Experimental Results and Analysis}
The aim of this section is to evaluate the efficacy of the MB-SVD model on real-world datasets. Extensive experiments and  empirical analysis are performed to address the ensuing inquiries:
\begin{itemize}
    \item \textbf{RQ1:} How does the performance of MB-SVD on various empirical datasets compared to the state of the art baselines?
   \item \textbf{RQ2:} How does the multi-behavior contrastive learning improve the efficiency of recommendation system?
    \item \textbf{RQ3:} How do varying configurations of hyperparameters impact the quality of the proposed model?
\end{itemize}

\begin{table}[htbp]
\caption{\rm{Statistical exploration of experimental datasets.}}
\centering
\scriptsize
\scalebox{1}{
\begin{tabular}{c|c|c|c|c}
\hline
    Dataset  & User & Item  & Interaction & Behavior Type \\ 
    \hline
Beibei  & 21716     & 7977      & $3.36 \times 10^6$        & \{View, Cart, Favorite,Purchase\}    \\
Taobao & 48749     & 39493      &  $2.0 \times 10^6$                    &         \{Click, Cart, Favorite,Purchase\}                        \\
Yelp & 19800     & 22734      &  $1.4 \times 10^6$                    &         \{Tip, Dislike, Neutral, Like\}                        \\
\hline
\end{tabular}}
\label{table1}
\end{table}

\begin{table*}
\caption{\rm{The performance of models on the Beibei, Taobao, and Yelp datasets is evaluated using the Recall@K and NDCG@K metrics with $K$ values of 10, 40, and 80.}
}
\centering
\scriptsize
\scalebox{1.1}{
\begin{tabular}{|c|c|c|c|c|c|c|c|c|c|c|c|c|c|c|}
\hline

\multicolumn{2}{|c|}{}& \multicolumn{4}{|c|}{Single-Behavior Models}& \multicolumn{5}{|c|}{Multi-Behavior Models}& \multicolumn{1}{|c|}{Our Model}\\
 \hline
Dataset& Metric & NCF & NGCF&ENMF&LightGCN&NMTR&EHCF&RGCN&MB-GMN&S-MBRec&MB-SVD\\
 \hline
\multirow{6}{4em}{Beibei} &Recall@10 &0.0251& 0.0389 &0.0377 &0.0452& 0.0462 &0.0459& 0.0480 &0.0497 &\underline{0.0529}&\textbf{0.0546}\\
&Recall@40& 0.0554& 0.0754& 0.0633 &0.1211&0.1366& 0.1271& 0.1263& 0.1498& \underline{0.1647}&\textbf{0.1652}\\
&Recall@80& 0.0641& 0.0933 &0.0812 &0.1939 &0.1992& 0.1923& 0.1912& 0.2017& \underline{0.2740}&\textbf{0.2820}\\
&NDCG@10 &0.0117 &0.0121& 0.0109& 0.0127 &0.0129 &0.0134& 0.0123 &0.0151& \textbf{0.0148}&\underline{0.0147}\\
&NDCG@40& 0.0164 &0.0154 &0.0171 &0.0187& 0.0193& 0.0214 &0.0226& 0.0397 &\underline{0.0429}&\textbf{0.0432}\\
&NDCG@80& 0.0228 &0.0206& 0.0312& 0.0334 &0.0423 &0.0439& 0.0443 &0.0465& \underline{0.0615}&\textbf{0.0618}\\
\hline
\multirow{6}{4em}{Taobao} &Recall@10 &0.0141& 0.0219& 0.0198& 0.3177& 0.0369 &0.0295 &0.0372 &0.0438 &\underline{0.0608}&\textbf{0.0611}\\
&Recall@40& 0.0204& 0.0297 &0.0224& 0.0405& 0.0487& 0.0599 &0.0706& 0.0873&\underline{ 0.1027}&\textbf{0.1041}\\
&Recall@80 &0.0311& 0.0763 &0.0459& 0.0795& 0.0983& 0.1030& 0.1527 &0.1559&\underline{ 0.1647}&\textbf{0.1664}\\
&NDCG@10& 0.0094& 0.0105& 0.0129 &0.0216 &0.0237 &0.0284 &0.0214& 0.0326& \underline{0.0391}&\textbf{0.0397}\\
&NDCG@40 &0.0141& 0.0162& 0.0226& 0.0287& 0.0305 &0.0374 &0.0304 &0.0398& \underline{0.0464}&\textbf{0.0470}\\
&NDCG@80 &0.0196& 0.0206& 0.0248& 0.0265& 0.0336 &0.0390& 0.0448& 0.0476& \underline{0.0583}&\textbf{0.0589}\\
\hline
\multirow{6}{4em}{Yelp} &Recall@10& 0.0114 &0.0175 &0.0163& 0.0148 &0.0197 &0.0186& 0.0205& 0.0243& \underline{0.0259}&\textbf{0.0262}\\
&Recall@40 &0.0375 &0.0398& 0.0407& 0.0676& 0.0724 &0.0705 &0.0843 &0.0879& \underline{0.1135}&\textbf{0.1142}\\
&Recall@80 &0.0498& 0.0604& 0.0535& 0.0823 &0.0634& 0.0980& 0.1090& 0.1398& \underline{0.1548}&\textbf{0.1554}\\
&NDCG@10 &0.0044& 0.0095 &0.0102 &0.0178 &0.0190 &0.0164& 0.0214& 0.0273& \underline{0.0287}&\textbf{0.0294}\\
&NDCG@40& 0.0141& 0.0162& 0.0126& 0.0187 &0.0305 &0.0294 &0.0204 &0.0248 &\underline{0.0337}&\textbf{0.0341}\\
&NDCG@80& 0.0164& 0.0216& 0.0227& 0.0235& 0.0354 &0.0342 &0.0398 &0.0416 &\underline{0.0438}&\textbf{0.0442}\\
\hline
\end{tabular}}\label{comparison}
\end{table*}

\subsection{Dataset}
To investigate the performance of our proposed MB-SVD model, we conduct extensive experiments on three real-world datasets, such as \textbf{Beibei} \citep{xia2021graph}, \textbf{Taobao} \citep{xia2021graph} and \textbf{Yelp} \citep{xia2021knowledge}. The statistical attributes of the datasets are summarized below, 
\begin{itemize}
 \item \textbf{Beibei.}
This dataset encompasses four distinct behaviors, namely \emph{view}, \emph{favorite}, \emph{add-to-cart} and \emph{purchase}. These behaviors represent different types of interactions that users have with items on the Beibei platform.

 \item \textbf{Taobao.}
This dataset comprises four distinct behaviors, namely \emph{click}, \emph{favorite}, \emph{add-to-cart} and \emph{purchase}. These behaviors represent different types of interactions that users have with items on the Taobao platform.
\item \textbf{Yelp.}
This dataset encompasses four distinct behaviors, namely \emph{tip}, \emph{dislike}, \emph{neutral} and \emph{like}. These behaviors represent different types of interactions of users towards businesses.
\end{itemize}

Similar to the study conducted by \citep{gu2022self}, we partition the datasets into distinct sets for training, validation, and testing purposes. As the datasets encompass a minimum of five discrete target behavior associations, we employ a stochastic selection procedure to allocate two of these associations, designating one as the principal testing set and the other as the designated validation set. The residual parts are employed for the purpose of training the model. Table \ref{table1} presents the comprehensive statistical information of the three datasets.

\subsection{Baselines and Evaluation Metrics}
We conduct a comparative analysis of our proposed model against nine baseline models. The baseline models are classified into two types: single-behavior models and multi-behavior models. The single-behavior models are as following:
\begin{itemize}
    \item \textbf{NCF}\citep{he2017neural} 
    : This framework for collaborative filtering is based on neural network principles, wherein a neural architecture is employed to supplant the inner product, thereby facilitating the acquisition of an arbitrary function from data.
    \item \textbf{NGCF} \citep{wang2019neural}
    : Leveraging the inherent graph configuration of user-item relationships, the recommendation model utilizes a graph convolution network to propagate embeddings and extract meaningful features.
    \item \textbf{ENMF} \citep{chen2020efficient}
    : It is a recommendation model that combines the power of neural networks and matrix factorization techniques. It is designed to learn user and item embeddings in an efficient manner by leveraging the interconnected user-item interaction graph.
    \item \textbf{LightGCN} \citep{he2020lightgcn}: It leverage the concept of neighborhood aggregation in GCNs to capture collaborative filtering signals effectively..
\end{itemize}

The chosen multi-behavior models include:
\begin{itemize}
    \item \textbf{NMTR} \citep{gutmann2010noise}: It leverages the power of multi-behavior data and optimizes collaborative filtering performance. 
    \item \textbf{EHCF} \citep{chen2020b}:  It is designed to capture intricate user-item relationships in a heterogeneous dataset and provide comprehensive recommendations without the use of negative sampling.
    \item \textbf{RGCN} \citep{schlichtkrull2018modeling}: It models the multi-relational data as a graph and applies convolutional filters on the graph structure to capture relational information. 
    \item \textbf{MB-GMN} \citep{xia2021graph}: It incorporates a graph-based matching network that leverages multiple behavior data to learn user and item embeddings, subsequently matching them within a structured graph framework.
    \item \textbf{S-MBRec} \citep{gu2022self}: It is a multi-behavior recommendation system that leverages a dual-task approach to model the diversity and similarity of multiple behaviors.
\end{itemize}

When evaluating the efficiency of our model, two commonly used evaluation metrics in the recommendation domain are adopted, that's Recall@K and NDCG @K\citep{krichene2020sampled}.

\subsection{Experimental settings}
We implement the MB-SVD model with the Pytorch framework. The experiments are conducted on a server with Intel(R) Core(TM) i5-7200U CPU @ 2.50GHz 2.71 GHz, 8.00 GB RAM, and an Intel(R) HD Graphics 620. The Adam optimizer is used to refine the MB-SVD model with a learning rate of $1e-4$. The model is trained with batch sizes selected from the set \{1024, 2048, 4096, 
6114\}, while the embedding dimension is explored within the range of \{64, 128, 256, 512\}. The task weight parameter $\lambda$ is probed within the interval of \{0.05, 0.1, 0.2, 0.5, 1.0\}, and $L_2$ regularization coefficient is chosen from the set of \{0.05, 0.1, 0.2, 0.5, 1.0\}. The temperature coefficient $\tau$ is probed within the set \{0.1, 0.2, 0.5, 1.0\}.

\subsection{Main Results(RQ1)}
Table \ref{comparison} presents the comparison results of our proposed MB-SVD model against a range of  baseline methods on three distinct datasets. To comprehensively evaluate the experimental results, we vary the parameter $N$ across \{10, 40, 80\}. The optimal results are presented in \textbf{Bold} font, while the subsequent best results are emphasized with \underline{underline}. Our analysis yields the following insights:

First, our MB-SVD model consistently and obviously surpasses alternative approaches across all datasets. In particular, on the  Taobao and Yelp datasets, our method demonstrates a remarkable improvement comparing with the most advanced benchmark methods. Moreover, our model achieves comparable performance on the Beibei dataset. 
These findings demonstrate the excellent performance of our model in both single-behavior and multi-behavior recommendation tasks. This highlights the model's remarkable proficiency and its ability to handle different recommendation scenarios.
By employing SVD decomposition, MB-SVD is able to mitigate the over-smoothing problem problem and retain important information for accurate recommendations.

Second, the multi-behavior models outperform single-behavior models, indicating that incorporating multiple behaviors enriches the semantics and improves behavior prediction accuracy. In particular, our proposed framework effectively captures and utilizes the diverse behaviors of users, resulting in improved recommendation accuracy and performance.
In addition, the augmentation mechanism of embedded representations and contrastive learning effectively handle the challenge of sparse data, leading to enhanced performance. 
\afterpage{
\begin{figure}[htbp]
 \begin{minipage}[t]{0.5\textwidth}
    \centering
    \includegraphics[width=0.8\linewidth]{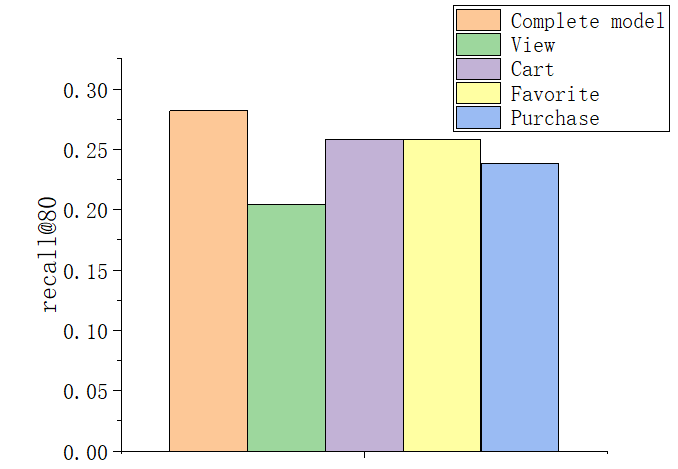}
    \centerline{(a) Beibei}
  \end{minipage}%
  
   \vspace{1cm} 
  \begin{minipage}[t]{0.5\textwidth}
    \centering
    \includegraphics[width=0.8\linewidth]{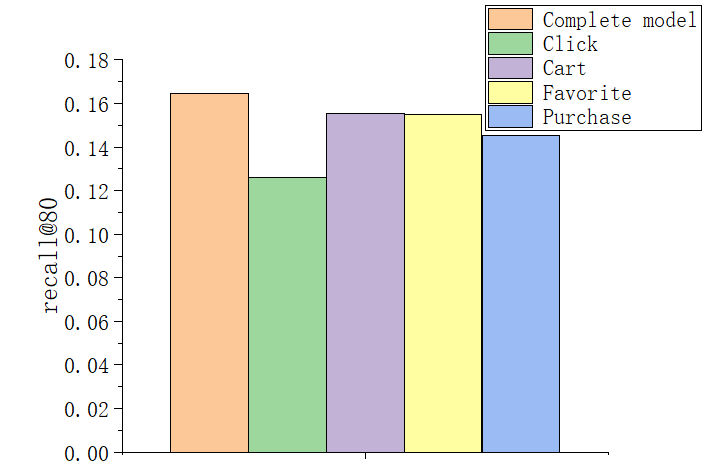}
    \centerline{(b) Taobao}
    \end{minipage}%

     \vspace{1cm} 
  \begin{minipage}[t]{0.5\textwidth}
    \centering
    \includegraphics[width=0.8\linewidth]{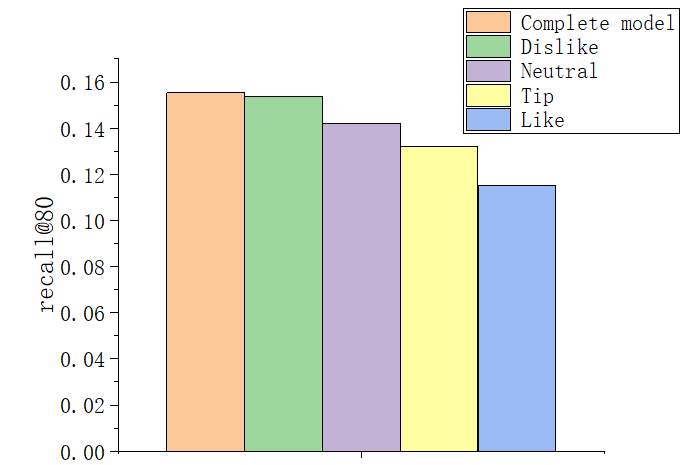}
    \centerline{(c) Yelp}
    \end{minipage}%
  \caption{\rm{Ablation study on various behaviors.}}
  \label{a1}
\end{figure}

\begin{figure}[htbp]
\centering
    \includegraphics[width=0.8\linewidth]{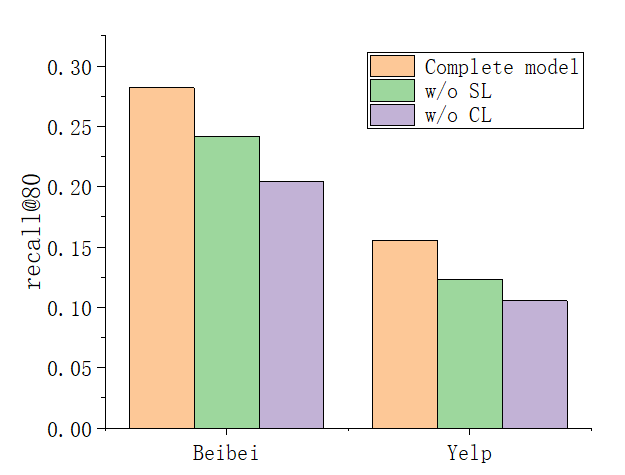}
      
    \caption{\rm{Ablation study on loss function $L_r$ and $L_s$}} \label{a2}
    \end{figure}
    
}

\subsection{Ablation Study(RQ2)}
We conduct ablation studies to validate the effectiveness of each behavior to our proposed model MB-SVD. For Beibei and Taobao datasets, we name the model without different behaviors as follows.
\begin{itemize}
    \item \textbf{View}: remove the \emph{View} behavior from the weighted multi-behavior representation.
    \item \textbf{Cart}: remove the \emph{cart} behavior from the weighted multi-behavior representation..
    \item \textbf{Favorite}: remove the \emph{Favorite} behavior from the weighted multi-behavior representation.
    \item \textbf{Purchase}: remove the \emph{Purchase} behavior from the weighted multi-behavior representation.
\end{itemize}

For Yelp dataset, we name the model without different behaviors as follows.
\begin{itemize}
    \item \textbf{Dislike}: remove the \emph{Dislike} behavior from the weighted multi-behavior representation.
    \item \textbf{Neutral}: remove the \emph{Neutral} behavior from the weighted multi-behavior representation..
    \item \textbf{Tip}: remove the \emph{Tip} behavior from the weighted multi-behavior representation.
    \item \textbf{Like}: remove the \emph{Like} behavior from the weighted multi-behavior representation.
\end{itemize}

Figure \ref{a1} shows the ablation results.
It is observed that our Complete model that includes all behaviors in datasets achieves the best results, which invalidates the efficiency and effectiveness of our multi-behavior weighted mechanism. Moreover, the influence of each behavior for the recommendation performance is different.
Among these behaviors, such as \emph{Favorite} and \emph{Cart} in the Taobao and Beibei datasets, their influence is remarkably similar and plays the most important role. The \emph{Dislike} behavior in the Yelp dataset has less impact compared to other behaviors. We can assign weight for each behavior according to their influence degree to the whole recommendation performance.

To better verify the role of loss function $L_r$ and $L_s$ (including $L_{us}$ and $L_{vs}$), and demonstrate the performance of joint optimization, we further design the following two variants of MB-SVD.

\begin{itemize}
    \item \textbf{w/o SL:} remove the  BPR loss function $L_r$ from the main objective function.
    \item \textbf{w/o CL:} remove the contrastive loss $L_s$ from the main objective function.
\end{itemize}

We compare the performance of two variants with the MB-SVD model on the the Beibei and Yelp datasets. Figure \ref{a2} shows the experimental results. It's interesting to note that omitting any term of the overall objective function results in a significant degradation of experimental outcomes. This observation suggests that both terms play a significant role in improving the performance of the model.
In addition, contrastive learning plays more important role in performance improvement. 
By combining these two loss functions in a joint optimization process, our model makes full use of the strengths of both approaches, and  improve the performance in terms of accuracy and recommendation quality.

\begin{figure}[htbp]
  \begin{minipage}[t]{0.5\textwidth}
    \centering
    \includegraphics[width=0.83\linewidth]{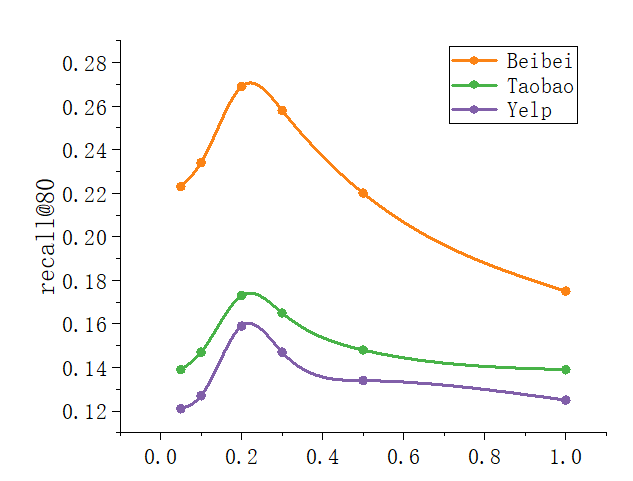}
    \centerline{(a) Impact of $\lambda$}
  \end{minipage}%
   \vspace{1cm} 
  \begin{minipage}[t]{0.5\textwidth}
    \centering
    \includegraphics[width=0.83\linewidth]{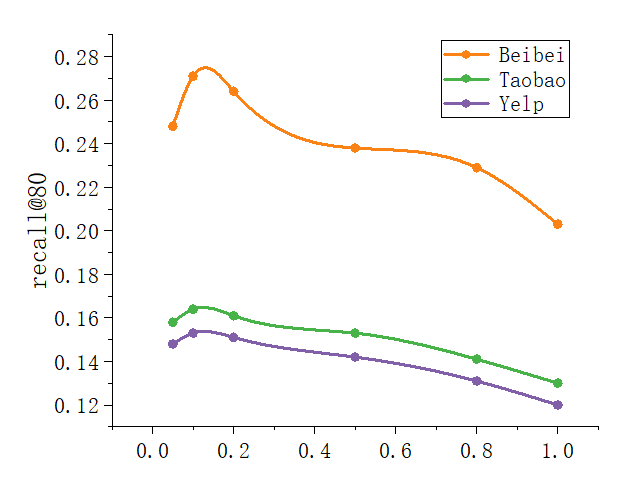}
    \centerline{(b) Impact of $\tau$}
    \end{minipage}%
     \vspace{1cm} 
    \begin{minipage}[t]{0.5\textwidth}
    \centering
    \includegraphics[width=0.83\linewidth]{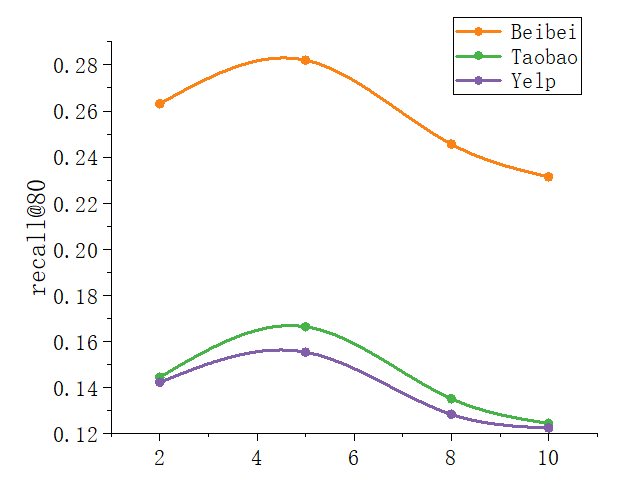}
    \centerline{(c) Impact of $q$}
    \end{minipage}%
  \caption{Hyperparameter effects of MB-SVD on three datasets.}
  \label{a3}
\end{figure}

\subsection{Hyperparameters Sensitivity(RQ3)}
In this section, we explore the effect of several key hyperparameters to the performance of the model, including  regularization weights for InfoNCE loss ($\lambda$), the temperature ($\tau$), and the required rank for SVD ($q$). Figure \ref{a3} visualizes the experimental results.

\begin{itemize}
    \item \textbf{The influence of $\lambda$:} We tune $\lambda$ in the range of \{0.05, 0.1, 0.2, 0.3, 0.5, 1.0\}, and evaluate the corresponding results. It is observed that increasing $\lambda$ can bring better results, however when it
    is greater than a certain point 0.2, the performance degrades obviously. It implies that the proportion of InfoNCE loss tends to be optimal with this setting. 
    \item \textbf{The influence of $\tau$:}  We tune $\tau$ in the range of \{0.05, 0.1, 0.2, 0.5, 0.8, 1.0\} and study the results. When it is greater than 0.2, increasing $\tau$ can make the performance decrease. The reason is that a small $\tau$ sharpens the similarity, while a greater $\tau$ smooths it. Therefore, $\tau$ = 0.2 ensures a suitable degree of similarity smoothness.
    \item \textbf{The influence of $q$:} The rank of SVD in MB-SVD is determined by the parameter $q$. From Figure \ref{a3}, we observe that the best performance is obtained when  $q$ is set to 5. The reason is that the model  maintains essential structure features within the interaction graph with this setting.
\end{itemize}

\section{Conclusion}
In this paper, we propose a novel MB-SVD framework for multi-behavior recommendation. 
In particular, our model designs an adversarial and simplified contrastive learning paradigm over both the SVD-augmentation representations and the original user and item representations. This approach allows the model to capture the important semantic information distilled through SVD augmentation while also considering the global collaborative context.
Furthermore, we integrate multi-behavior learning tasks to enrich the embedded representations with more informative features and effectively solve the problem of data sparsity. Our experimental results on various datasets demonstrate the superior performance of our method.
In future work, we plan to investigate the potential of
incorporating learnable data augmentation and distillation technique into our graph contrastive learning model to further improve the performance of the recommendation system. By employing learnable data augmentation, we can adaptively generate augmented views for the dataset, potentially resulting in more informative representations. Furthermore, the use of distillation techniques can help incorporate knowledge from external sources and improve the robustness and effectiveness of the recommendation model.

\section{Acknowledgements}
\par This work was supported by the National Natural Science Foundation of China under Grant No.62172243 and the China Postdoctoral Science Foundation under Grant No.2022M711088.

\bibliographystyle{apalike}
\bibliography{ref.bib}

\end{document}